\documentclass{ptapap}

\author{Varadarajan Parthasarathy}[CAMK]
\author{Antonios Manousakis}[CAMK]
\author{W{\l}odzimierz Klu{\'z}niak}[CAMK]
\affil[CAMK]{Nicolaus Copernicus Astronomical Center\\
  Bartycka 18, 00--716 Warszawa, Poland}

\title{Numerical simulations of oscillating tori in a pseudo-Newtonian potential}

\begin{document}

\maketitle

\begin{abstract}

We have modeled hydrodynamical tori in the Klu{\'z}niak-Lee pseudo-Newtonian potential. The tori in 
equilibrium were perturbed with uniform sub-sonic velocity fields: vertical, radial and diagonal respectively, 
and allowed to evolve in time. We identify the eigenmodes corresponding to those of slender tori. The results of our simulations are relevant to the investigation of high-frequency quasi-periodic oscillations observed in stellar-mass
black hole binaries. 

\end{abstract}

\section{Introduction}

Several interesting astrophysical phenomena are observed in accretion flows in the strong field limit, from regions close to compact bodies like black holes or neutron stars. In this regard the X-ray binaries (the central compact body is a low-magnetic-field neutron star or a stellar-mass black hole) are astrophysical laboratories for observing phenomena in a regime where Newtonian physics will not be valid. The Rossi X-ray Timing Explorer (RXTE) satellite launched on December 30, 1995, successfully observed several phenomena predicted during the 70's and 80's. In 1996 RXTE discovered the signals of kilohertz quasi-periodic oscillation (QPO) which were predicted by \citep{1973SvA....16..941S}, caused by orbital motion near the innermost stable circular orbit from general relativistic effects. The motivation for investigating QPOs are manifold, like understanding the dynamics of the inner regions of accretion disks, determining the mass and the spin of the central compact body, testing general relativity, etc \citep{2005ASIB..210..283V}. The study of QPOs has deep impact on understanding disk seismology \citep{2001PASJ...53....1K,1999PhR...311..259W}. 

We have performed axisymmetric hydrodynamical simulations of perturbed tori using the numerical code PLUTO\footnote{Freely available at http://plutocode.ph.unito.it/} \citep{2007ApJS..170..228M}. The tori were modeled in the pseudo-Newtonian potential obtained by \citep[hereafter KL]{2002MNRAS.335L..29K}
\begin{equation}
\Phi_{\rm KL} = -\frac{GM}{3r_{\rm g}}\left(e^{3r_{\rm g}/R} - 1\right) 
\end{equation}
where $r_{\rm g} = 2GM/c^{2}$ and $R = \sqrt{r^{2} + z^{2}}$ is the distance from the compact object. The innermost stable 
circular orbit (ISCO) is at 3$r_{\rm g}$. The KL potential reproduces the ratio of orbital and epicyclic frequencies for a Schwarszchild black hole:
\begin{equation}
\omega_{\rm r} = \Omega_{\rm K} \sqrt{1 - \left(\frac{3r_{\rm g}}{r}\right)} \qquad ,
\end{equation}
where $\omega_{\rm r}$ is the radial epicyclic frequency and $\Omega_{\rm K}$ is the orbital frequency. 

The equipotential surfaces of equilibrium polytropic torus corresponding to constant distribution of angular momentum 
\citep{1978A&A....63..221A,2000ApJ...528..462H} take the form
\begin{equation}
\Phi_{\rm KL} + \frac{1}{2} \frac{\ell_{\rm c}^{2}}{r^{2}} + \frac{\gamma}{\gamma - 1} \frac{P}{\rho} = C 
\end{equation}
with $P$ = constant, where $P = K\rho^{\gamma}$, $K$ denotes the polytropic constant, and we take $\gamma = 5/3$. The constant of integration $C$ is determined from the condition $P = 0$. The parameter space investigated and the frequencies of various modes obtained from the simulations are described in our recent work \citep[submitted to MNRAS]{2015arXiv151102663P}.

\section{Simulations}
We studied \textit{thinner} tori (cross-sectional radii $r_{\rm t} = 0.18 r_{\rm g}$) and a \textit{thicker} torus (cross-sectional radius $r_{\rm t} = 0.36 r_{\rm g}$). The results obtained from the vertical oscillation of the thicker torus are discussed. The torus in equilibrium is perturbed vertically and allowed to evolve in time. The L2-norm of the density ($\left\Vert \rho \right\Vert_{2}$) and the corresponding power density spectra (PDS) obtained from the vertical oscillation of thicker torus are displayed in Fig. 1. The modes excited in the vertical oscillation of thicker torus are: vertical ($V_{\rm h}$), breathing (B) and x-mode. The frequencies of various modes obtained from the simulations agree with those of slender tori \citep{2006MNRAS.369.1235B}. The radial mode (R) and the
plus mode (+), also observed in the simulations are numerical artefacts. Indeed, the simulation of unperturbed thicker torus excited the radial, plus and breathing modes. The breathing mode is manifest in the entire parameter space of the simulations, since the equilibrium torus will not be in perfect equilibrium due to inherent numerical truncation error and interaction of torus 
with the background. 

The vertical oscillation of an infinitely slender torus is tantamount to that of a rigid body \citep{2005AN....326..820K, 2004ApJ...617L..45B}. The eigenfrequency is equal to the vertical epicyclic frequency ($\omega_{\rm z}$), which for spherically symmetric metrics is equal to the orbital frequency ($\Omega_{\rm K}$). However the eigenmode of the simulated torus is more complicated. The vertical oscillation of the thicker torus exhibited a \textit{see-saw} motion due to its non-slenderness ($t = 462$, $t = 490$ and $t = 518$ in Fig. 2). We observe such a see-saw motion at the vertical epicyclic frequency, hence our simulation was successful in teasing out the actual eigenmode of vertical oscillation.  As the torus is symmetric with respect to the equatorial plane, we inferred the power of the vertical mode and the x-mode to be concentrated in the corresponding first harmonic respectively (right panel Fig. 1). 

Our motivation to add velocity perturbations to observe peaks of QPOs are similar to a recent work \citep{2015arXiv151007414M} on general relativistic hydrodynamical simulations of relativistic slender tori. The eigenfrequencies identified from the simulations of oscillating relativistic tori have correspondence to those identified from our simulations. 

\begin{figure}
\label{fig:big}
\includegraphics[width=\textwidth]{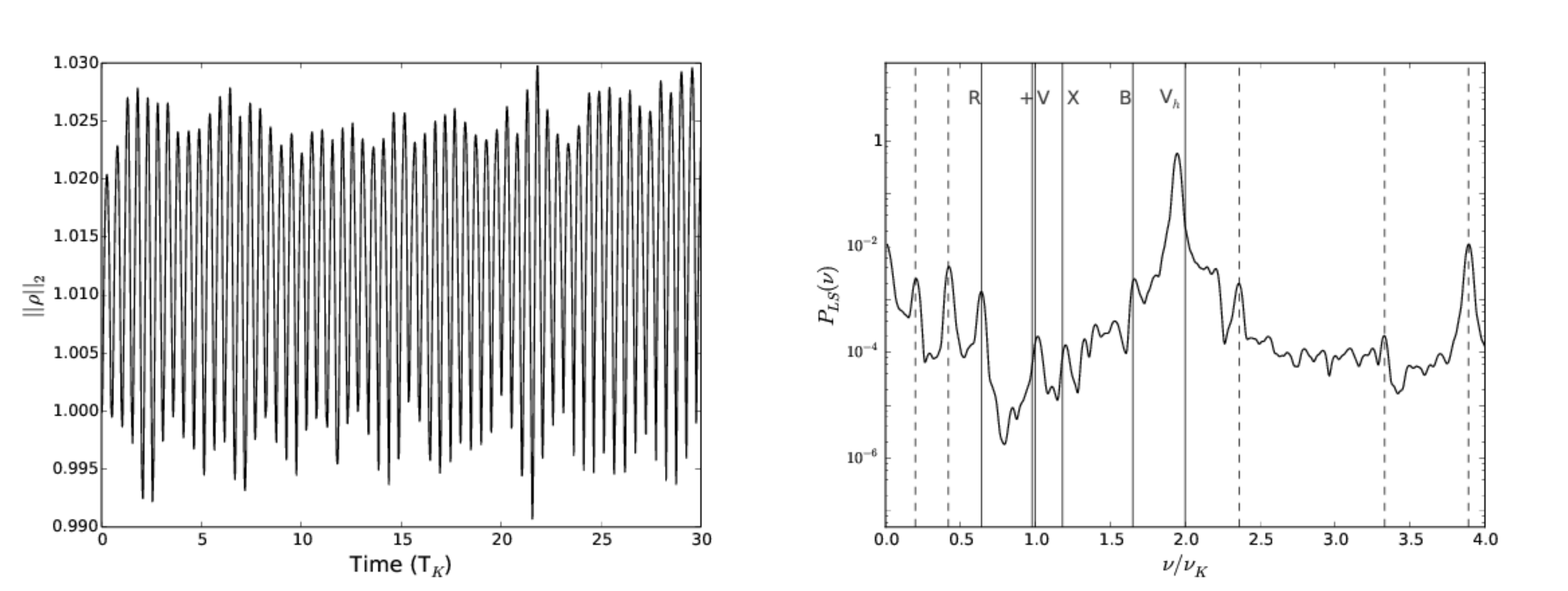}
\caption{\emph{Left}: $\left\Vert \rho \right\Vert_{2}$ as a function of time in units of the Keplerian period ($T_{\rm K}$) at $r_{\rm c} = 5.2r_{\rm g}$ for the thicker torus. \emph{Right:} The corresponding  PDS of $\left\Vert \rho \right\Vert_{2}$. Frequencies are in units of Keplerian frequency ($\nu_{\rm K} = 1/T_{\rm K}$ at $r_{\rm c} = 5.2r_{\rm g}$). The solid lines correspond to the theoretical values of fundamental eigenfrequencies of slender tori modes and dashed lines correspond to harmonics or additional frequencies apparently present in the simulations.}
\end{figure}

\begin{figure}
\label{fig:v46}
\includegraphics[width=\textwidth]{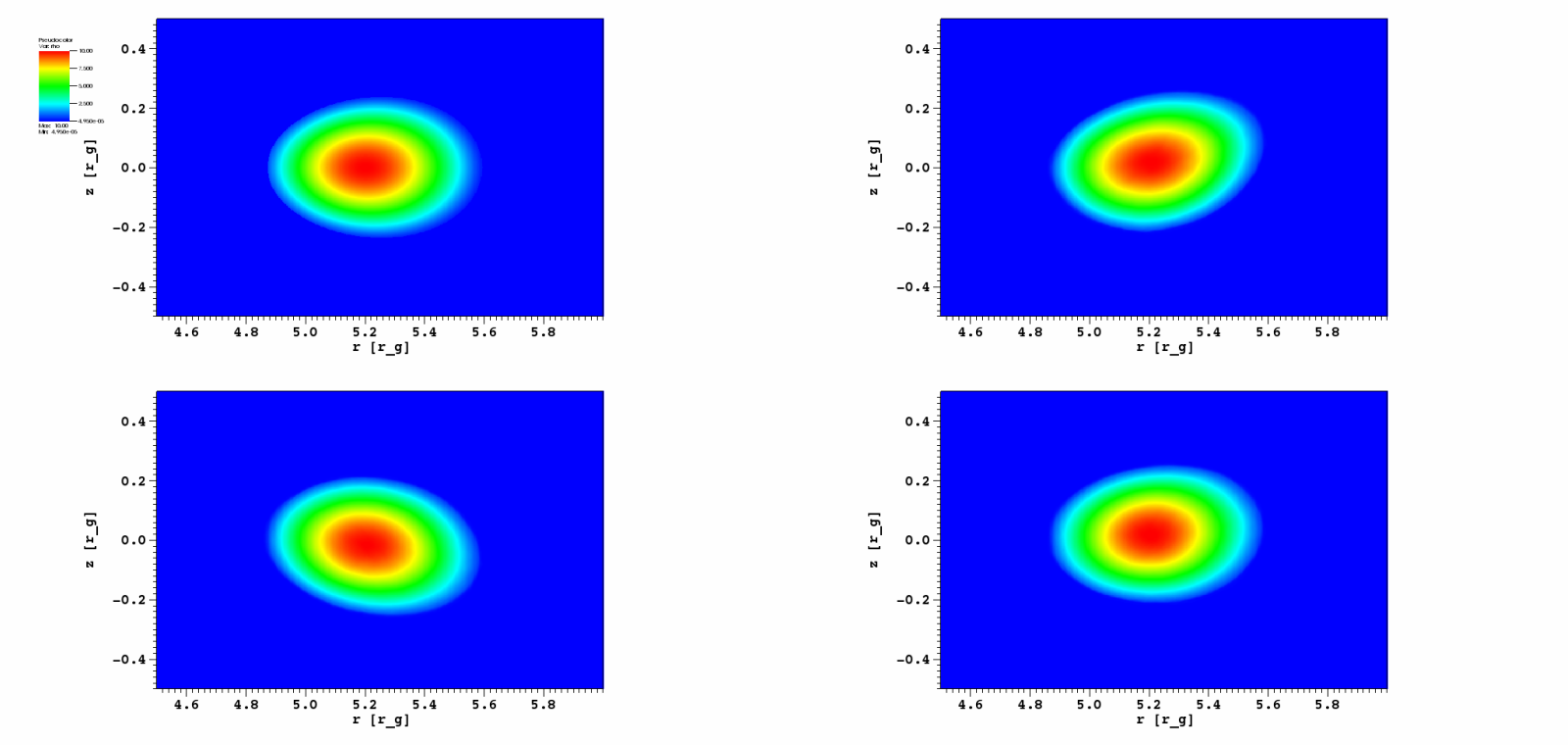}
\caption{Density of vertically oscillating torus. The center of the torus ($r_{\rm c}$) is at 5.2$r_{\rm g}$. Clock-wise from the top left are snapshots of the pseudo-color plots of mass density at $t = 0$, $t = 462$, $t = 490$ and $t = 518$ respectively. Units of time are such that $T_{\rm K}$ = 56.}
\end{figure}

\section{Conclusions}

We have performed axisymmetric hydrodynamical simulations of perturbed tori in the Klu{\'z}niak-Lee potential and investigated the excited modes by adding uniform sub-sonic velocity perturbations to the tori in equilibrium. We have found that the vertical eigenmode includes both an overall vertical oscillation and a superimposed see-saw motion of the cross-section of the torus. The parameter space of the simulations included the thinner tori and the thicker torus. The vertical oscillation of thicker torus exhibits the breathing mode (B), x-mode and vertical mode ($V_{\rm h}$). The radial mode and the plus mode excited in the vertical oscillation are numerical artefacts --- the simulation of unperturbed thicker torus exhibited the radial and plus modes. The results of our simulations correspond to those obtained in the Schwarszchild metric. 

\acknowledgements{Our research was supported by Polish NCN grant 2013/08/A/ST9/00795.}

\bibliographystyle{ptapap}
\bibliography{ptapapdoc}

\begin{thebibliography}{13}
\providecommand{\natexlab}[1]{#1}
\providecommand{\url}[1]{\texttt{#1}}
\providecommand{\urlprefix}{URL }
\providecommand{\eprint}[2][]{\url{#2}}

\bibitem[{{Abramowicz} et~al.(1978){Abramowicz}, {Jaroszynski}, \&
  {Sikora}}]{1978A&A....63..221A}
{Abramowicz}, M., {Jaroszynski}, M., {Sikora}, M., \emph{{Relativistic,
  accreting disks}}, \emph{\aap} \textbf{63}, 221 (1978)

\bibitem[{{Blaes} et~al.(2006){Blaes}, {Arras}, \&
  {Fragile}}]{2006MNRAS.369.1235B}
{Blaes}, O.~M., {Arras}, P., {Fragile}, P.~C., \emph{{Oscillation modes of
  relativistic slender tori}}, \emph{\mnras} \textbf{369}, 1235 (2006),
  \eprint{astro-ph/0601379}

\bibitem[{{Bursa} et~al.(2004){Bursa}, {Abramowicz}, {Karas}, \&
  {Klu{\'z}niak}}]{2004ApJ...617L..45B}
{Bursa}, M., {Abramowicz}, M.~A., {Karas}, V., {Klu{\'z}niak}, W., \emph{{The
  Upper Kilohertz Quasi-periodic Oscillation: A Gravitationally Lensed Vertical
  Oscillation}}, \emph{\apjl} \textbf{617}, L45 (2004),
  \eprint{astro-ph/0406586}

\bibitem[{{Hawley}(2000)}]{2000ApJ...528..462H}
{Hawley}, J.~F., \emph{{Global Magnetohydrodynamical Simulations of Accretion
  Tori}}, \emph{\apj} \textbf{528}, 462 (2000), \eprint{astro-ph/9907385}

\bibitem[{{Kato}(2001)}]{2001PASJ...53....1K}
{Kato}, S., \emph{{Basic Properties of Thin-Disk Oscillations}}, \emph{\pasj}
  \textbf{53}, 1 (2001)

\bibitem[{{Klu{\'z}niak}(2005)}]{2005AN....326..820K}
{Klu{\'z}niak}, W., \emph{{High frequency QPOs, nonlinear oscillations in
  strong gravity}}, \emph{Astronomische Nachrichten} \textbf{326}, 820 (2005),
  \eprint{astro-ph/0510725}

\bibitem[{{Klu{\'z}niak} \& {Lee}(2002)}]{2002MNRAS.335L..29K}
{Klu{\'z}niak}, W., {Lee}, W.~H., \emph{{The swallowing of a quark star by a
  black hole}}, \emph{\mnras} \textbf{335}, L29 (2002),
  \eprint{astro-ph/0206511}

\bibitem[{{Mignone} et~al.(2007)}]{2007ApJS..170..228M}
{Mignone}, A., et~al., \emph{{PLUTO: A Numerical Code for Computational
  Astrophysics}}, \emph{\apjs} \textbf{170}, 228 (2007),
  \eprint{astro-ph/0701854}

\bibitem[{{Mishra} et~al.(2015)}]{2015arXiv151007414M}
{Mishra}, B., et~al., \emph{{Quasi-periodic oscillations from relativistic
  hydrodynamical slender tori}}, \emph{ArXiv e-prints}  (2015),
  \eprint{1510.07414}

\bibitem[{{Parthasarathy} et~al.(2015){Parthasarathy}, {Manousakis}, \&
  {Kluzniak}}]{2015arXiv151102663P}
{Parthasarathy}, V., {Manousakis}, A., {Kluzniak}, W., \emph{{Quasi-periodic
  oscillations of perturbed tori}}, \emph{ArXiv e-prints}  (2015),
  \eprint{1511.02663}

\bibitem[{{Syunyaev}(1973)}]{1973SvA....16..941S}
{Syunyaev}, R.~A., \emph{{Variability of X Rays from Black Holes with Accretion
  Disks.}}, \emph{\sovast} \textbf{16}, 941 (1973)

\bibitem[{{van der Klis}(2005)}]{2005ASIB..210..283V}
{van der Klis}, M., \emph{{Timing Neutron Stars}}, in A.~{Baykal}, S.~K.
  {Yerli}, S.~C. {Inam}, S.~{Grebenev} (eds.) NATO Advanced Science Institutes
  (ASI) Series B, \emph{NATO Advanced Science Institutes (ASI) Series B},
  volume 210, 283 (2005)

\bibitem[{{Wagoner}(1999)}]{1999PhR...311..259W}
{Wagoner}, R.~V., \emph{{Relativistic diskoseismology.}}, \emph{\physrep}
  \textbf{311}, 259 (1999), \eprint{astro-ph/9805028}

\end{thebibliography}

\end{document}